\documentclass[preprint,aps,floats,showpacs,nofootinbib]{revtex4}
\usepackage{graphicx}
\usepackage{epsfig}
\usepackage{pstricks}
\usepackage{pst-coil}
\usepackage{amsmath}
\usepackage{verbatim}

\def\be{\begin{equation}}
\def\ee{\end{equation}}
\def\bea{\begin{eqnarray}}
\def\eea{\end{eqnarray}}
{
{
\def\mpl{M_{\rm {Pl}}}
\def\gev{{\rm \,Ge\kern-0.125em V}}
\def\tev{{\rm \,Te\kern-0.125em V}}
\def\mev{{\rm \,Me\kern-0.125em V}}

\def\Tmax{T_{\rm {max}}}
\def\Treh{T_{\rm {reh}}}
\def\Tf{T_{\rm {f}}}
\def\T1{T_1}
\def\tmax{t_{\rm {max}}}
\def\treh{t_{\rm {reh}}}
\def\Hreh{H_{\rm {reh}}}
\def\Hmax{H_{\rm {max}}}
\def\Rmax{R_{\rm {max}}}
\def\nG{n_{\tilde G}}
\def\YG{Y_{\tilde G}}
\def\mG{m_{\tilde G}}
\def\rhoG{\rho_{\tilde G}}
\def\greh{g_{*\rm{reh}}}
\def\Sigmatot{\Sigma_{{\rm {tot}}}}

\begin{document}
\title{\bf Gravitino production in an inflationary Universe and 
implications for leptogenesis}
\author{Raghavan Rangarajan}
\email{raghavan@prl.res.in}
\author{Narendra Sahu}
\email{narendra@prl.res.in}
\affiliation{Theoretical Physics Division, Physical Research Laboratory,
Navarangpura, Ahmedabad 380 009, India}
\begin{abstract}
\noindent
Models of leptogenesis are constrained by the low reheat 
temperature at the end of reheating associated with the gravitino 
bound. However a detailed view of reheating, in which the maximum 
temperature during reheating, $\Tmax$, can be orders of magnitude 
higher than the reheat temperature, allows for the production of 
heavy Majorana neutrinos needed for leptogenesis. But then one must 
also consider the possibility of enhanced gravitino production in 
such scenarios. In this article we consider gravitino production 
during reheating, its dependence on $\Tmax$, and its relevance for 
leptogenesis. Earlier analytical studies of the gravitino abundance
have only considered gravitino production in the post-reheating 
radiation dominated era. We find that the gravitino abundance 
generated during reheating is comparable to that generated
after reheating. This lowers the upper bound on the reheat temperature by
a factor of 4/3.
\end{abstract}
\maketitle
\section{Introduction}
\noindent
Gravitinos in supersymmetric theories can have important cosmological 
consequences. Stable gravitinos can overclose the universe while unstable
gravitinos can affect the expansion rate of the universe during eras 
prior to their decay. The decay products of unstable gravitinos
can also overclose the universe or affect light element abundances 
generated during nucleosynthesis. These cosmological consequences are 
a function of the gravitino energy density, $\rhoG=\mG\nG$, where 
$\mG$ and $\nG$ are the mass and number density of gravitinos.

In a non-inflationary universe, $\nG\sim T^3$ and therefore cosmological 
constraints on the energy density of gravitinos provide bounds on 
$\mG$, and equivalently on the scale of supersymmetry breaking~\cite{
pagelsprimack,weinberg}. In an inflationary universe, $\nG$ is also a 
function of the reheat temperature, and so for a fixed $\mG$, often 
taken to be $O(100\gev-1\tev)$, cosmological constraints on the
energy density of gravitinos provide an upper bound on the reheat 
temperature~\cite{nos,krauss,khlopovlinde,EKN,falomkinetal,jss,ENS,ks,
kletal,moroi.95,cefo}.

In Refs.~\cite{nos,krauss,khlopovlinde,EKN,falomkinetal,jss,ENS,ks,
kletal,moroi.95,cefo} the number density of gravitinos is obtained by 
considering gravitino production in the radiation dominated era 
following reheating. It is presumed that $n_{\tilde G}=0$ at the 
beginning of the radiation dominated era. Gravitinos are then produced 
through thermal scattering and $n_{\tilde G}$ is found to be proportional 
to the reheat temperature, $\Treh$, which is the temperature of the
thermal plasma at the beginning of the radiation dominated era
when the inflaton field has decayed completely and the energy density 
of the universe is dominated by the inflaton decay products. The 
cosmological constraints on $\nG$ then provide an upper bound on 
$\Treh$ of $10^{6-9}$\gev. 

If, as in Refs.~\cite{nos,krauss,khlopovlinde,EKN,falomkinetal,jss,ENS,
ks,kletal,moroi.95,cefo}, one assumes instantaneous reheating after 
inflation or that $\Treh$ is the maximum temperature during reheating, 
then the upper bound on $\Treh$ makes it difficult to create 
sufficiently high number densities of GUT gauge and Higgs bosons whose 
decays could generate the baryon asymmetry of the universe. Similarly, 
the gravitino bound constrains leptogenesis models in which
the lepton asymmetry is generated by the decay of heavy bosons
or fermions~\cite{delepinesarkar}. However, as discussed in Refs.~\cite{
kolb_book,chungetal}, after the inflationary era the temperature 
does not rise instantaneously to $\Treh$ but rises initially to a 
maximum temperature $\Tmax$ and then falls to $\Treh$.  
In Ref.~\cite{chungetal} the authors then argue that $\Tmax$ can be as 
high as $10^3 \Treh$ and that sufficient numbers of the lightest heavy 
Majorana right-handed neutrino of mass $\sim 10\Treh$ can be produced 
during reheating to allow for successful leptogenesis. This issue has 
also been studied in Ref.~\cite{delepinesarkar,giudiceetal}.

While the above scenario considers the possible production of heavy 
neutrinos during reheating it does not consider the possible enhancement 
in the gravitino production as well. If the gravitino abundance 
generated post-reheating is proportional to the maximum temperature 
during that era, namely $\Treh$, one should ask whether the gravitino 
abundance generated during reheating is proportional to the maximum 
temperature, $\Tmax$, of the reheating era.  If this abundance is larger
than the abundance generated in the post-reheating era it could affect
the viability of the leptogenesis scenario of Ref.~\cite{chungetal}. 
Therefore in this article we explicitly calculate the gravitino abundance 
generated during the reheating era.  We then compare it to the standard 
calculation of the gravitino abundance generated after reheating in the 
radiation dominated era, and discuss its relevance for leptogenesis models. 

As argued above, one might expect that the gravitino abundance generated 
during reheating will be a function of $\Tmax$. Interestingly, what we find
is that by manipulating the relations between $\Tmax$, $\Treh$, the inflaton
decay rate and the scale of inflation the dependence on $\Tmax$ cancels out
and the gravitino abundance is proportional to $\Treh$ only. Furthermore, 
while one would not expect the gravitino abundance generated in the 
reheating and the post-reheating eras to be similar that is indeed what we 
find. The resulting contraint on the reheat temperature and on
leptogenesis models is then only slightly stronger than before.

Our results are valid for a reheating scenario that does not include 
preheating.  Gravitino production during preheating has been considered
in Ref.~\cite{greene.99,maroto.00,kallosh.00,tsujikawa.00,nilles.01,
nilles&olive.01,podolsky}.

\section{Production of gravitinos} 
During inflation the Universe cools down by several orders of magnitude.  
Subsequently the inflaton decays while performing coherent oscillations 
about the minimum of its potential. Very soon after the inflaton enters 
the oscillating phase the temperature of the universe rises to a maximum 
value~\cite{kolb_book} 
\be
T_{\rm {max}}\simeq 0.8g_*^{-1/4}M_{{\rm I}}^{1/2}
\left( \Gamma_\phi \mpl\right)^{1/4}\,,
\label{T-max}
\ee
where $M_{{\rm I}}=V_{{\rm I}}^{1/4}$, $V_{{\rm I}}$ being the vacuum 
energy density during the inflationary epoch (taken to be constant). 
$g_*$ is the number of relativistic degrees of freedom and $\Gamma_\phi$ 
is the decay rate of the inflaton field. Subsequently, the temperature 
of the thermal bath falls approximately as $R^{-3/8}$~\cite{kolb_book}, 
where $R$ is the scale factor of expansion of the Universe. This particular 
dependence on $R$ goes on until the universe becomes radiation dominated when 
the inflaton field decays completely at $t_{\rm{reh}}=\Gamma_\phi^{-1}$.
The temperature of the universe at $\treh$ is given by \cite{kolb_book} 
\be
\Treh\simeq 0.55 \greh^{-1/4}\left( \mpl\Gamma_\phi \right)^{1/2}\,.
\label{reheat-temp}
\ee
In the following we examine the production of gravitinos during reheating 
between $\Tmax$ and $\Treh$, and during the subsequent radiation 
dominated era, and discuss its consequences.

Gravitinos are produced by the scattering of the inflaton decay products; 
see, for example, Tables 1 in Refs.~\cite{EKN,moroi.95} for a list
of processes. The number density of gravitinos generated is then given by 
the solution of the Boltzmann equation 
\be
\frac{dn_{\tilde{G}}}{dt}+3Hn_{\tilde{G}}=
\langle \Sigmatot|v|\rangle n^2\,,
\label{boltzmann}
\ee
where $n=(\zeta(3)/\pi^2)T^3$ ($\zeta(3)=1.20206..$ is the Riemann zeta 
function of 3), $\Sigmatot$ is the total scattering cross section for 
gravitino production, $v$ is the relative velocity of the incoming
particles, and $\langle...\rangle$ refers to thermal averaging. We have 
ignored the gravitino decay term above as the gravitino lifetime is 
$~10^{7-8} (100\gev/\mG)$s~\cite{EKN} and is not relevant during the
gravitino production era for gravitinos of mass $10^{2-3}\gev$. 
We may re-express this equation as
\be
\dot T\frac{dn_{\tilde{G}}}{dT}+3Hn_{\tilde{G}}=\langle \Sigmatot|v|\rangle
n^2\,,
\label{boltzmann1}
\ee
In $SU(N)$ supersymmetric models with $n_f$ pairs of fundamental and
antifundamental chiral supermultiplets, $\langle\Sigmatot|v|\rangle$
is given by~\cite{BBB_npb.01}
\bea
\langle\Sigmatot|v|\rangle\equiv
\frac{\alpha}{M^2}
&=& \frac{1}{M^2} \left[ 1+\left( \frac{m_{\tilde{g}}^2}{3\mG^2}\right)\right]
\frac{3g_N^2(N^2-1)\pi}{32\zeta (3)}\nonumber\\
&\times& [\{\ln(T^2/m_{g,\rm{th}}^2)+0.3224\}(N+n_f)+0.5781n_f]\,,
\label{sigma-total}
\eea
where $m_{\tilde{g}}$ is the gaugino mass and $m_{g,\rm{th}}$ is the thermal
mass of the gauge boson which is given as
\be
m_{g,\rm{th}}^2=\frac{1}{6}g_N^2(N+n_f)T^2\,.
\ee
In the above equations $g_N\,(N=1,2,3)$ are the gauge coupling constants
corresponding to $U(1)_Y$, $SU(2)_L$ and $SU(3)_C$ respectively and $M=\mpl/
\sqrt{8\pi}\simeq 2.4\times 10^{18}$ GeV is the reduced Planck mass.
(For $U(1)$ gauge interactions, $N^2-1\rightarrow 1$ and $N+n_f\rightarrow 
n_f$, where $n_f$ is the sum of the square of the hypercharges of chiral 
multiplets~\cite{kkm}.) Using the one loop $\beta$-function of MSSM, the 
solution of the renormalization group equation for the gauge coupling 
constants is given by
\be
g_N(T)\simeq \left[ g_N^{-2}(M_Z)-\frac{b_N}{8\pi^2}\ln(T/M_Z)
\right]^{-1/2}\,,
\label{coupling-const}
\ee
with $b_1=11$, $b_2=1$, $b_3=-3$.
It is presumed here that inflaton decays perturbatively and the
products thermalise quickly as discussed in Appendix A of Ref.~\cite{
chungetal}. Also see Refs.~\cite{mazumdar1,mazumdar2} for an alternate 
description of reheating and the gravitino bound.  

\subsection{Gravitino production during reheating}
If the potential for the oscillating inflaton field is dominated by
the mass term, then the energy density of the inflaton field scales 
as $1/R^3$ during reheating. (We ignore change in the inflaton energy 
density due to decays.) Einstein's equation then implies
\be
\left (\frac{\dot{R}}{R} \right)^2 = \frac{8\pi G}{3}\rho_{\rm max}
\left( \frac{R_{\rm max}}{R} \right)^3\,,
\label{H-rho-relation}
\ee
where $\rho_{\rm max}$ and $R_{\rm max}$ are the inflaton energy density 
and scale factor at $\Tmax$. Solving the above equation for $R$ we get
\be
R=R_{\rm max}\left[ {3\over 2}\Hmax(t-\tmax)+1\right]^{2/3}\,,
\label{R-matter-phase}
\ee
where 
\be
\Hmax\simeq \sqrt{ \frac{8\pi}{3} }\frac{M_I^2}{\mpl}\,.
\label{H-max-value} 
\ee
(For $t\gg t_{\rm max}$, $R\sim t^{2/3}$.)

For Eq. ({\ref{boltzmann1}}) we require $\dot T$ and $H$ as functions of 
the temperature $T$. During reheating the energy density of the relativistic 
particles is given by $\rho_r\propto R^{-3/2}$~\cite{kolb_book}.  
Since $\rho_r\sim T^4$, we get 
\be
T=\Tmax\left( \frac{\Rmax}{R} \right)^{3/8}\,.
\label{R-T-relation}
\ee
Using Eq. (\ref{R-matter-phase}) the $T-t$ relation is then obtained as
\be
T={\Tmax}\frac{1}{\left[\frac{3}{2}\Hmax(t-\tmax)+1\right]^\frac{1}{4}}\,.
\label{T-t}\ee
This implies that 
\begin{eqnarray}
\dot T &=&-\frac{3}{8}T \Hmax\frac{1}{\left[\frac{3}{2}\Hmax(t-\tmax)+1\right]}
\nonumber\\
&=&-\frac{3}{8}T \Hmax\left( \frac{T}{\Tmax}\right)^4 \,.
\label{Tdot}\end{eqnarray}
The Hubble expansion parameter in terms of $T$ is then given by
\begin{equation}
H=\frac{\dot{R}}{R}=-\frac{8}{3}\frac{\dot{T}}{T}=\Hmax\left( 
\frac{T}{\Tmax} \right)^4\,.
\label{H-T-relation}
\end{equation}

Using Eqs. (\ref{Tdot}) and (\ref{H-T-relation}), Eq. (\ref{boltzmann1}) 
can be rewritten as 
\be
\frac{dn_{\tilde{G}}}{dT}-\frac{8}{T} n_{\tilde{G}}=-CT\,,
\label{final-boltzmann}
\ee
where 
\be
C=\frac{8}{3}\frac{\Tmax^4}{\Hmax}\frac{\alpha}{M^2} 
\left(\frac{\zeta(3)}{\pi^2}\right)^2\,.
\label{c-value}
\ee
Solving for $n_{\tilde{G}}$ in the regime $\Tmax$ to $\Treh$ we get 
\begin{eqnarray}
n_{{\tilde{G}}}(\Treh)&=&\frac{C}{6}\Treh^8
\left(\frac{1}{\Treh^6}-\frac{1}{\Tmax^6}\right)
\label{boltz-solution-a}\\
&=&\frac{C}
{6}\Treh^2\quad \rm{for \,\,\Tmax\gg\Treh} \,.
\label{boltz-solution}
\end{eqnarray}
$\alpha$ has been taken to be constant although there is a $\log$ 
dependence due to the running gauge couplings in the one loop $\beta$ 
function of MSSM. 

\subsection{Gravitino production in the radiation dominated era}
After the inflaton field decays completely at $\treh$ the universe 
enters the radiation dominated era. Unlike the reheating era during 
which the entropy continuously increases, in the radiation dominated 
era the total entropy remain constant (except for epochs of 
out-of-equilibrium decays). Therefore it is useful to express the 
abundance of any species $i$ as $Y_i=n_i/s$, where $n_i$ is 
the number density of the species $i$ in a physical volume 
and $s$ is the entropy density given by
\be
s=\frac{2\pi^2}{45}g_*T^3\,.
\label{entropy-density}
\ee
$g_*=228.75$ in the MSSM. With this definition, Eq. (\ref{boltzmann1}) 
reads as
\be
\dot T\frac{d\YG}{dT}=\langle \Sigmatot|v|\rangle
 Y n \,.
\label{boltzmann2}
\ee
For the radiation dominated era,
\be
T={\Treh}\frac{1}{\left[2\Hreh(t-\treh)+1\right]^\frac{1}{2}}\,,
\label{T-t-reh}\ee
where 
\be
\Hreh= \sqrt{ \frac {8\pi^3 g_{*\rm{reh}}} {90} }\frac{\Treh^2}{\mpl}\,.
\label{H-reh-value} 
\ee
This implies that $\dot T$ is
\begin{equation}
\dot T =-\frac{\Hreh}{\Treh^2}T^3 = 
-\left( \frac{g_{*\rm{reh}}\pi^2}{90}\right)^{\frac{1}{2}}\frac{T^3}{M}\,.
\end{equation}
Then
\be
\frac{d\YG}{dT}=-\left( \frac{90}{\greh\pi^2} \right)^{1/2}
\left(\frac{1}{(2\pi^2/45)g_*}\right) \left(\frac{\alpha}{M}\right)
\left( \frac{\zeta(3)}{\pi^2} \right)^2\,.
\label{YT-boltzmann}
\ee
Assuming $\alpha$ to be independent of temperature and integrating 
the above equation from $\Treh$ to $\Tf$, the final temperature, 
we get the number density of gravitinos at $\Tf$ to be
\bea
Y_{\tilde{G}}(\Tf) &=& Y_{\tilde{G}}(\Treh)+\left( \frac{90}{\greh\pi^2} 
\right)^{1/2}\left(\frac{1}{(2\pi^2/45)\greh}\right)\nonumber\\
&&\times  \left(
\frac{\alpha}{M}\right)\left( \frac{\zeta(3)}{\pi^2} \right)^2 (\Treh-\Tf)\,.
\label{gravitino-density}
\eea
Since most of the gravitinos are generated close to $\Treh$ we have 
ignored the variation of $g_*$ with temperature and used $\greh$ in
the final expression. Using Eqs. (\ref{boltz-solution}), (\ref{c-value}) 
and (\ref{entropy-density}) the first term on the right hand side of the 
above equation is given by
\be
Y_{\tilde{G}}(\Treh)=\frac{\alpha}{M^2}\left( \frac{\zeta(3)}
{\pi^2} \right)^2\left(\frac{1}{(2\pi^2/45)\greh}\right) \frac{4}{9}
\frac{\Tmax^4}{\Hmax\Treh}\,. 
\label{density-treh}
\ee
This term is usually neglected while estimating the gravitino 
abundance~\cite{nos,krauss,khlopovlinde,EKN,falomkinetal,jss,ENS,
ks,kletal,moroi.95,cefo}. However as we see below this is comparable 
with the second term in Eq. (\ref{gravitino-density}). Now using 
Eq. (\ref{density-treh}) in Eq. (\ref{gravitino-density}) we get the 
effective number density of gravitinos at $\Tf$ to be
\bea
Y_{\tilde{G}}(\Tf) &=& \frac{\alpha}{M^2}
\left( \frac{\zeta(3)}{\pi^2} \right)^2
\left(\frac{1}{(2\pi^2/45)\greh}\right) \left[\frac{4}{9}\frac{\Tmax^4}
{\Hmax\Treh}\right.\nonumber\\
&&+ \left. M\left( \frac{90}{\greh\pi^2} \right)^{1/2} \Treh\right]\,,
\label{comparison}
\eea
where we have used $\Tf\ll\Treh$. Relating $\Tmax$ to $\Treh$ from Eqs. 
(\ref{T-max}) and (\ref{reheat-temp}) the number density of gravitinos 
in a comoving volume is then given as
\be
Y_{\tilde{G}}(T_f)=\frac{\alpha \Treh}{M}
\left( \frac{\zeta(3)}{\pi^2} \right)^2\left( \frac{1}
{(2\pi^2/45)\greh^{3/2}}\right)(1.0+3.0)\,,
\label{comparison-value}
\ee
where we have used $\greh$ in the expressions for $\Tmax$.

\section{Discussion and comments}
\begin{figure}[t]
\begin{center}
\epsfig{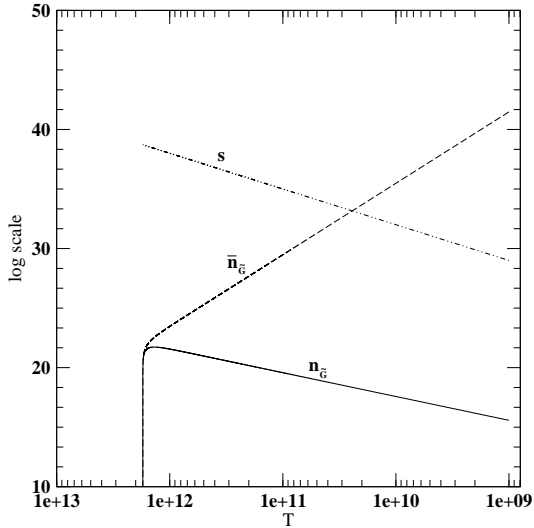}
\end{center}
\caption
{\small\sf The gravitino number density, $\nG$, the gravitino number
density per comoving volume, ${\bar n}_{\tilde G}$, 
and the entropy density, $s$, are plotted
as a function of the temperature $T$ during reheating.  $\Treh$ and $M_I$
are chosen
to be $10^9\gev$ and $10^{16}\gev$ respectively, 
and so $\Tmax\approx2\times10^{12}\gev$.  $\T1\approx\Tmax$.
$\alpha$ is treated as constant and evaluated at $\Treh$, 
with $g_i(M_z)$ obtained from
$\alpha_{EM}(M_Z)=1/128$, $\sin^2\theta_W(M_Z)=0.231$, $\alpha_s(M_Z)=0.119$,
and $M_Z=91.2\gev$ \cite{pdg}.
}
\label{fig1}
\end{figure}

\begin{figure}
\begin{center}
\epsfig{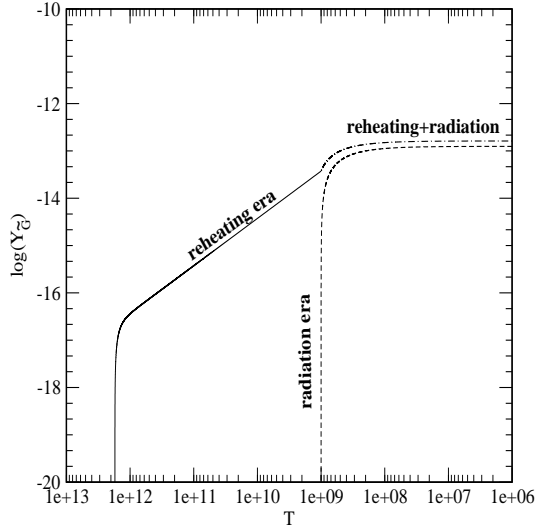}
\end{center}
\caption[]
{\small\sf $\YG=\nG/s$ generated during the reheating era, 
the radiation dominated
era, and the sum of contributions from both eras
are shown as a function of the temperature $T$ for the same parameters
as in Fig. (\ref{fig1}).
Since $\YG$ in both eras is largely 
generated close to $\Treh$,
$\alpha$ is evaluated at $\Treh$.
The final value of $\YG$ is $\approx 2\times 10^{-13}$.
}
\label{fig2}
\end{figure}

Generalizing Eq. (\ref{boltz-solution-a}) for any $T$ during reheating 
one sees that $\nG$ does not vary monotonically during reheating.  
As shown in Fig. (\ref{fig1}), $\nG$ rises dramatically
from $\Tmax$ to $\T1=\Tmax/4^{1/6}$ and then falls from $\T1$ to 
$\Treh$. However $R^{3}\sim T^{-8}$ and the number density per 
comoving volume, ${\bar n}_{\tilde G}=(1/\Rmax^3)\int_{\Tmax}^T dT\, 
d/dT(\nG R^3)$, is proportional to $(1/T^6-1/\Tmax^6)$ and so most 
gravitinos are produced close to $T\sim \Treh$. ($s\sim T^3$ and, as shown in Fig. (\ref{fig2}), $\YG$ 
increases steadily in the reheating phase for $\Treh<T<\Tmax$.)
For $T<\Treh$, one can show from Eq. (\ref{gravitino-density}) that
$d {n}_{\tilde G}/dT$ is always greater than 0, indicating that the gravitino
number density is always decreasing during the radiation dominated era.
(If $\YG(\Treh)$ is set to 0, then $\nG$ first increases and then decreases.)
$\YG$ is always increasing but for $T\ll\Treh$ it becomes approximately
constant, as seen in Fig. (\ref{fig2}).

We now address a concern as to whether one can distinguish
production as occuring in the reheating era or in the radiation dominated era 
when production
in both eras occurs close to $\Treh$.  In our analysis, for $T<\Tmax$
during reheating the
ratio of the radiation and inflaton energy densities increases as
$\rho_{{\rm rad}}/\rho_{{\rm inf}} \sim R^{-3/2}/R^{-3} \sim R^{\,3/2} 
\sim T^{-4}$ while the gravitino
abundance increases as $\YG \sim T^{-1}$ (from Eq. (\ref{density-treh})).  
Therefore much of the change in
the gravitino abundance in the reheating era occurs when 
$\rho_{{\rm rad}}\ll \rho_{{\rm inf}}$.
Similarly, Eq. (\ref{gravitino-density})
implies that in the radiation dominated era the gravitino
abundance changes linearly with $T \sim t^{-1/2}$, while the inflaton energy
density falls exponentially fast.  Again much of the gravitino production
at $T<\Treh$
will occur when it is valid to treat the dynamics of the era as due to
radiation only.  Therefore we believe that our treatment of the problem
is valid.  

%

From Eq. (\ref{comparison-value}) it is clear 
that the gravitino production during the reheating
period is 1/3 of that during the radiation dominated era.
A priori one would not have expected the gravitino production
in both these eras to be similar.  Furthermore, the gravitino abundance
generated during reheating can be re-expressed as independent of $\Tmax$,
which is also unexpected.
Interestingly, the contribution to $\YG$ from the reheating era is
linearly proportional
to $\Treh$, as it is for the 
radiation dominated era.
It is then straightforward to revise the earlier
upper bound on $\Treh$ 
from the cosmological constraints on $\nG$.
The bound
of $10^{6-9} \gev$ 
will now be lowered by a factor
of 4/3 and thus is not greatly affected.
Since $\Tmax\propto\sqrt\Treh$, $\Tmax$ is also not much affected.
Therefore heavy particles of mass greater than $\Treh$
can still be produced during reheating and leptogenesis scenarios
are not significantly further constrained than before.

{\bf Comparison with numerical analysis:} 
The total gravitino density, including that generated during reheating, 
has been obtained
numerically in Ref. \cite{kkm}.  Our analytical 
derivation agrees well with their 
fit to the gravitino abundance.  For a reheat temperature of
$10^9\gev$, both
Eq. (F12) of Ref. \cite{kkm} and our Eq. (\ref{comparison-value})
give $\YG=2\times10^{-13}$ indicating the robustness of our analysis.
Furthermore, our analytical derivation allows one to appreciate various
aspects of the gravitino abundance obtained in Ref. \cite{kkm}.
The fit to the gravitino abundance in 
Ref. \cite{kkm} gives an abundance dependent on $\Treh$, but not also 
on $\Tmax$ as one might have expected.  Our analysis above shows
that the abundance generated during reheating does
indeed depend on $\Tmax$ (see Eq. (\ref{density-treh})) but it also depends 
on the scale
of inflation and $\Treh$, and by manipulating the expressions relating 
these three
quantities, as we have done, the dependence on $\Tmax$ cancels out.

The dominant term in the fit of Ref. \cite{kkm} for the total
gravitino abundance has the same functional form as that obtained by
earlier analytical calculations that did not include gravitino generation during
reheating, namely, linear dependence on $\Treh$.  
Our analysis indicates that 
this is because
gravitino generation during reheating
also has a linear dependence on $\Treh$, just as in the post-reheating era. 
We emphasise that it would be improper to naively conclude that the 
linear dependence is because the gravitino
abundance generated during the radiation dominated era is dominant, since
we have shown that the gravitino abundance
generated in both eras differ only by a factor of 3.

\section{Conclusion}

In conclusion, in this article we have calculated the 
gravitino abundance generated during reheating.
We find that it is linearly
proportional to the the reheat temperature $\Treh$, as in the standard
calculation of gravitinos produced in the radiation dominated
era after reheating.
Further, we find that it is about 1/3 the number density of gravitinos 
generated in the radiation dominated era.
Therefore this lowers the upper bound on $\Treh$ from cosmological constraints
on the gravitino number density
by a factor of 4/3. This 
does not significantly 
alter the viability of leptogenesis scenarios.

%

\end{document}